\documentclass[12pt]{article}
\usepackage{graphicx}
\newcommand{\grs}{GRS~1915+105~}

\begin{document}
\begin{figure}[t]
\begin{center}
\includegraphics[angle=0,width=1.0\textwidth ]{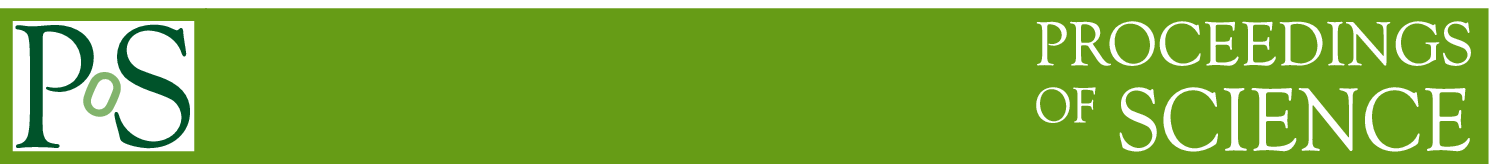}
\end{center}
\end{figure}

\Large
\centerline{\bf Origin of Superluminal radio jets in GRS 1915+105 and} 
\centerline{\bf the role of the Plateau state }
\large
\vskip 0.3in
\centerline{\bf J S Yadav}
\vskip 0.01in
\centerline{\bf Tata Institute of Fundamental Research, Homi Bhabha Road,}
\vskip 0.01in
\centerline{\bf Mumbai-400005, India}
\vskip 0.5in  

\begin{abstract}
{\bf We have studied the accretion disk during the radio plateau state  and the 
following superluminal relativistic radio jets and have provided a tight
correlation between accretion disk and superluminal jet parameters. We
find that accretion rate during the plateaux is very high and suggest 
that the accretion disk during the radio plateaux  is always associated with
radiation-driven  wind. The internal shock  forms in the previously generated 
slowly moving wind (during plateau)  with $\beta$ $\le$ 0.01  as the 
fast moving discrete jet (usually at the end of plateau) with $\beta$ $\sim$ 1
catches up and 
interacts with it.  The power of superluminal jet is  determined by  
the strength and  speed of these two components;  the slow moving wind and  
the fast moving jet which are related to the accretion disk during the  
plateau state. Finally, we discuss the  implication of this work.}
\end{abstract}
\vskip 1.5in
\noindent VI Microquasar Workshop: Microquasars and Beyond\\
\noindent		 September 18-22, 2006\\
\noindent		 Como, Italy
\newpage
\normalsize
\normalfont
\section{Introduction}

In active galactic nuclei (AGN) \& quasars, large superluminal outflows/jets  
are known to exist for almost half century
yet remain poorly understood. This is probably due to the lower characteristic
accretion disk temperature as well as the fact that the central part of 
these systems is  
usually obscured by a large amount of dust. The long time scales associated 
with these massive systems pause an additional observational problem. 
Discovery of the first microquasar \grs in 1994 in our Galaxy revived 
great hope to study the disk-jet connection in general and the origin 
of superluminal jets in particular
\cite{mira94}.
Microquasars  are 
closer, smaller and show faster variability that is easily observable.
 \grs  has shown exceptionally
high variability in  X-rays as well as in  radio. Since 1996, rich X-ray 
variability of this source is observed by RXTE \cite{morg97,muno99} and 
by the Indian X-ray Astronomy Experiment, IXAE \cite{yada99}.  Belloni et al.
(2000)
have classified the complex X-ray variability of \grs in 12 separate 
classes on 
the basis of their light curves and  color-color diagrams and suggested
three basic states of this source, namely  hard state and two   softer states 
with different temperatures of the accretion disk.

	The radio emission from \grs can be broadly put  into two 
classes; (1) radio emission close to the compact object (<200 AU), and
(2) radio emission at large distances ($\ge$ 240 AU).  The former 
class includes
(a) steady radio jets (radio plateau state), (b) preplateau flares,
and (c) oscillations/baby
jets (discrete jets) of 20-40 min duration in infrared (IR) \& radio
while the latter class includes
large superluminal  radio jets. The steady radio jets of 20 -- 160
mJy flux density are associated with the canonical low hard X-ray state
and observed for extended durations \cite{muno01,fuch03}. These are  
optically thick compact jets with  velocity $\beta$ of 0.1--0.4 
\cite{dhaw00,ribo04}. The radio emission is correlated  with the X-ray
emission as L$_{radio}$ $\propto$ L$_{X}^{0.7}$ for several different 
sources \cite{gall03}. Pooley \& Fender  (1997) observed  radio oscillations with
delayed emission at  lower frequency. Simultaneous X-ray, IR and radio 
multi-wavelength observations provided first major step in our understanding
of disk-jet interaction and suggested that the spike in X-ray  coincides 
with the beginning of IR flare and it has self absorbed synchrotron 
emission associated with adiabatic expansion \cite{eike98,mira98}.
These are also compact jets with velocity $\beta$ $\sim$ 1 
\cite{dhaw00}. The relativistic superluminal jets with up to 1 Jy flux density have steep radio
spectrum and are observed at large distances  few hundred AU to 5000 AU from
the core \cite{mira94,fend99,dhaw00}. These radio jets are very energetic with
luminosity close to the Eddington luminosity, L$_{Edd}$ and have been observed
in several sources \cite{fend99,hjel95,wu02,oros01}. Progress in our understanding
of these jets, especially of  their connection to  the accretion disk has 
been slow.   
The physical connection between X-ray emission  and the superluminal flares
  has been  the hardest to understand.

In this paper, we investigate the association of large superluminal 
jets with the 
radio plateaux. We have analysed the available RXTE PCA/HEXTE X-ray data
during radio plateaux and the radio flare data from the Green Bank 
Interferometer (GBI). Yadav (2006) has described the analysis procedure and 
the selection criteria in detail.
  
\begin{table}[h]
\begin{center}
\scriptsize
\begin{tabular}{|lccc|clccc|}
\hline
\hline
\multicolumn{4}{|c|}{Superluminal Radio Flare Properties }  &\multicolumn{5}{|c|}{Associated Preceding Plateau Properties} \\
\hline 
MJD/ & Peak&Rise&Decay Time  & GBI 
&Date of &QPO &Total & N$_H$ \\
Date &Flux & Time  &constant &Flux &  RXTE& Freq.&X-ray&(10$^{22}$ cm$^{-2}$) \\
 &(mJy) &(day) & (days)  & (mJy) &Obser.
&(Hz)&Flux$^a$& \\
\hline
50750/1997 Oct 30  &550 &$<$0.6 &3.16$^{+0.07}_{-0.09}$ &47.3  &1997 Oct. 25&1.88&2.04&13.27$^{+0.66}_{-0.91}$ \\
50916/1998 Apr 13  &920 &$<$0.8 &4.02$^{+0.36}_{-0.33}$ &91.0  &1998 Apr. 11&1.60&2.42&14.95$^{+0.82}_{-0.65}$ \\
50933/1998 Apr 30  &580 &$<$0.9 &3.98$^{+0.13}_{-0.13}$ &91.0  &1998 Apr. 28&1.41&2.34&13.55$^{+0.78}_{-0.31}$ \\
50967/1998 Jun 03  &710 &0.3    &2.82$^{+0.48}_{-0.28}$ &56.2  &1998 May 31 &1.76&2.14&12.99$^{+0.81}_{-0.32}$ \\
51204/1999 Jan 30  &340 &0.25   &1.12$^{+0.03}_{-0.05}$ &27.8  &1999 Jan. 24&2.55&1.84&11.19$^{+0.92}_{-0.95}$  \\
51337/1999 Jun 08  &490 &$<$0.7 &2.67$^{+0.22}_{-0.16}$ &45.5  &1999 Jun. 03&1.77&1.98&12.43$^{+0.95}_{-1.15}$  \\
51535/1999 Dec 23  &510 &$<$0.8 &2.67$^{+0.12}_{-0.14}$ &51.3  &1999 Dec. 21&2.12&2.23&13.61$^{+1.17}_{-0.94}$  \\
52105/2001 Jul 16$^b$  &210 &$<$0.7&1.77$^{+0.20}_{-0.20}$ &20.0  &2001 Jul. 11&2.40&1.75&10.60$^{+0.77}_{-0.40}$  \\
\hline
\multicolumn{9}{l}{{\bf a:} Integrated 3--150 keV X-ray flux in 10$^{-8}$ ergs cm$^{-2}$ s$^{-1}$} \\
\multicolumn{9}{l}{{\bf b:} VLBA radio data (see text for details)} \\
\end{tabular}
\caption{All selected superluminal radio  flares and their properties (see text for
details). X-ray Properties from RXTE PCA/HEXTE data during preceding
radio plateau.}
\label{radio_sftab}
\end{center}
\end{table}
\section{Observations and analysis}

\begin{figure}[t]
\begin{center}
\includegraphics[angle=270,width=.7\textwidth]{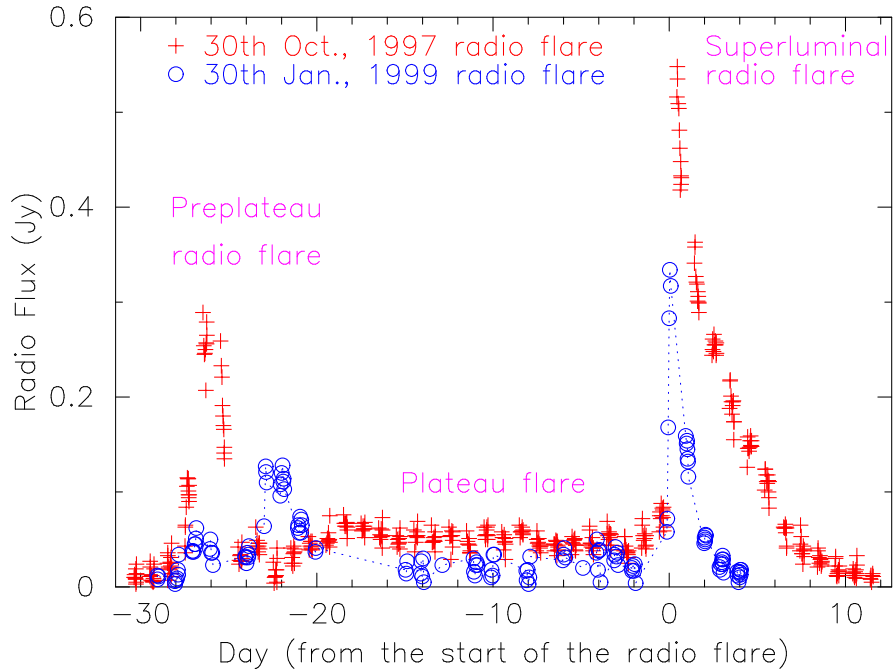}
\caption{GBI 2.25 GHz radio data for the 1999 January 30 (start at MJD 51204.7)
and 1997 October 30 (start at MJD 50750.6) large superluminal radio flares.
Start of the flares is offset to 0.  The dotted line
connects the data points of the 1999 January 30 superluminal radio flare to
guide the eyes.}
\label{morph}
\end{center}
\end{figure}

       The typical sequence of events for a superluminal radio flare is 
shown in  Figure~\protect\ref{morph} for the 550 mJy radio flare on 1997 October 30
(gap in GBI data at peak) and for the 340 mJy radio flare on 1999 January 30.
The start of the radio flares is offset to zero. A superluminal event starts 
with small preplateau flares followed by a steady long plateau, followed by
superluminal radio flares.  The preplateau flares  are  discrete
ejections of adiabatically expanding synchrotron clouds with flat radio 
emission and are similar to the oscillations/discrete jets.
The exponential decay is an important characteristic of superluminal
radio flares which differentiates them  from the other class of radio flares 
which occur 
close to the compact object \cite{yada06}. It is also found that the radio 
plateau is always associated with a superluminal
radio  flare \cite{fend99,klei02,vada03}. We searched 2.25 GHz GBI radio
monitoring data during the period from 1996 December to 2000 April and
selected radio plateaux and the following radio flares
which decay exponentially. All the selected superluminal flares are
given in Table~\protect\ref{radio_sftab} along with associated X-ray properties.
One more radio flare on 2001 July 16 which was observed by the Very Large 
Baseline Array (VLBA) and Ratan radio telescope is also 
added in Table~\protect\ref{radio_sftab}.
VLBA observations clearly showed an ejecta well separated 
from the core \cite{dhaw03}.

   We study X-ray properties during radio plateaux within the
preceding week from the start of superluminal flares to avoid changes in
accretion disk over the long durations  of plateaux (from -7 to -2
in Figure~\protect\ref{morph}).  We also avoid the last day of plateaux as radio data
suggest rapid changes in  the accretion disk. For the timing analysis,
we used single-bit-mode
RXTE/PCA data in the energy ranges 3.6 -- 5.7 and 5.7 -- 14.8
keV. Normalised power density spectra of 256 bin  are generated and
co-added for every 16 s.
During the plateaux, it  is
in a very high luminosity state (VHS) with power law index $\Gamma$ $>$ 2 .
The spectrum is dominated by the Compton scattered emission ($\ge$ 85\%)
rather than the disk.  A model consisting of disk +  power law + a comptonised 
component (CompTT) with a gaussian line at 6.4 keV is used 
for the X-ray analysis.
The integrated radio flux is calculated by integrating the fitted 
exponential function over a duration three times the decay time constant.
Further details of X-ray  and radio data analysis are given in Yadav (2006)
\cite{yada06}.

\section{Results and discussion}
It is increasingly  believed that the coronal material (and not the disk 
 material) is ejected prior to radio flares \cite{rau03, vada03,fend04,
 roth05}. It is 
 consistent with the observation that radio flares are observed during 
the  transition from the low hard state to the high soft state but never 
 observed during the opposite transition \cite{klei02}.
It also agrees  with the suggestion  that the spike in X-ray during
the $\beta$- class is 
associated  with the change of the X-ray state due to a major ejection
episode \cite{yada01}. Once this coronal material is ejected, it  
invariably decouples from the disk. However, this decoupling may not 
produce any observable effect during the plateaux as inflow and outflow in
the accretion disk   are in equilibrium.  Both the accretion disk and 
 the radio flare are in steady state. On the other hand, decoupling is
 supposed to produce maximum effect during the superluminal radio flares
 which are observed at large distances of a    
 few hundred to few thousand AU from the core \cite{muno01,fend99,dhaw00}. This has
 made disk-jet connection the hardest to understand for superluminal 
 radio jets \cite{fend04,fend04b}.
\begin{figure}[t]
\begin{center}
\includegraphics[angle=270,width=.6\textwidth]{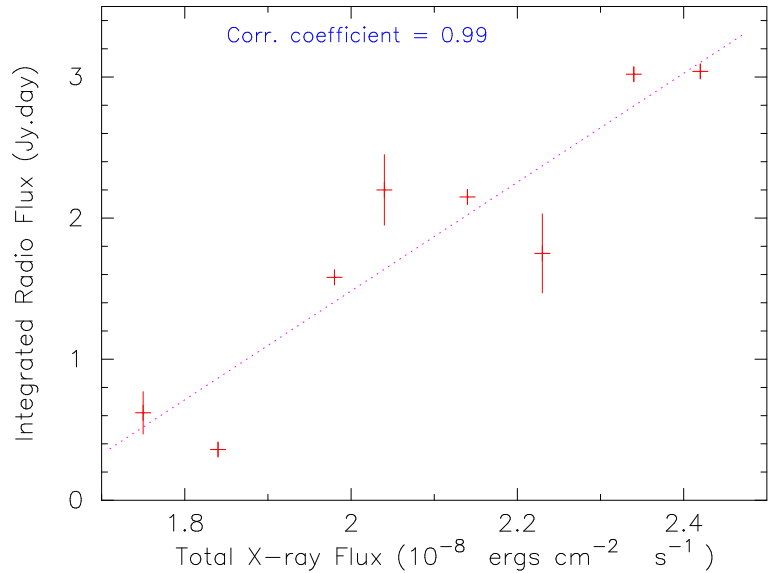}
\caption{Correlation between the total 3-150 keV X-ray flux
from spectral analysis
of RXTE PCA/HEXTE X-ray data during preflare plateau state
and the integrated flux of the superluminal radio flare. The solid line is the
linear fit to the data points.}
\label{xrayradio}
\end{center}
\end{figure}

\begin{figure}[t]
\begin{center}
\includegraphics[angle=270,width=.6\textwidth]{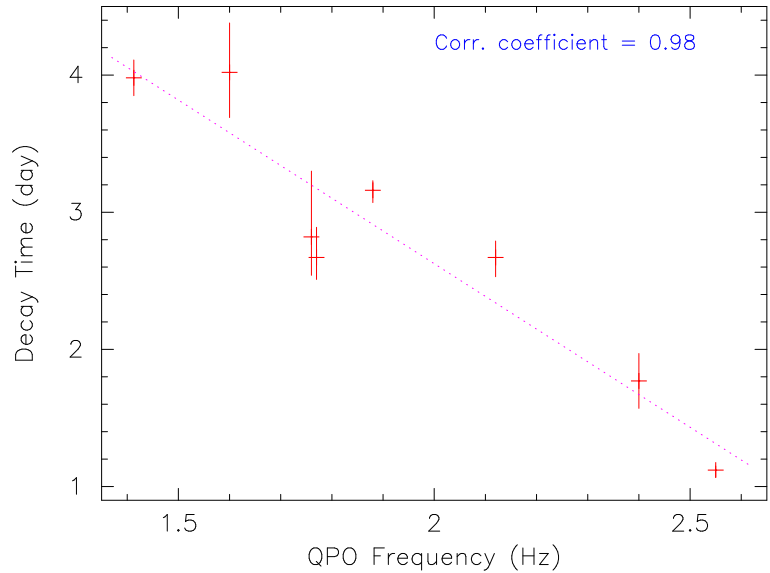}
\caption{Correlation between the quasi-periodic oscillation (QPO) frequency
from timing analysis of  RXTE/PCA
X-ray data during preflare plateau state and the decay time constant of
the superluminal radio flare. The dotted line is the linear fit to the data.}
\label{qpodt}
\end{center}
\end{figure}

In Figure~\protect\ref{xrayradio}, we plot the total 3--150 keV
 X-ray flux during the preceding plateau vs the integrated radio flux of 
 the superluminal flare. We derive a correlation coefficient of 0.99 suggesting 
a strong connection between the total X-ray flux during the preceding plateau 
state and the integrated radio 
flux of the superluminal flare. The total X-ray flux is dominated by the  
Compton scattered emission or coronal emission ($\ge$ 85\%). Using the
total Compton scattered emission flux instead of the total X-ray flux improves 
this correlation with a  correlation coefficient of 0.996.
In Figure~\protect\ref{qpodt}, 
we plot the 
QPO frequency from our timing analysis of X-ray data during the plateau state as
a function of the decay time constant of the following  
superluminal radio flare.
This also  shows a tight correlation, with correlation coefficient of 0.98. 
The QPOs are believed to be associated with the coronal flow.
The
remarkable feature of our findings here  is that the parameters calculated using
completely independent spectral and timing analysis  bring out a clear 
connection between the accretion disk during the plateau state  and the 
following superluminal radio flare.  

\begin{figure}[h]
\begin{center}
\includegraphics[angle=270,width=.7\textwidth]{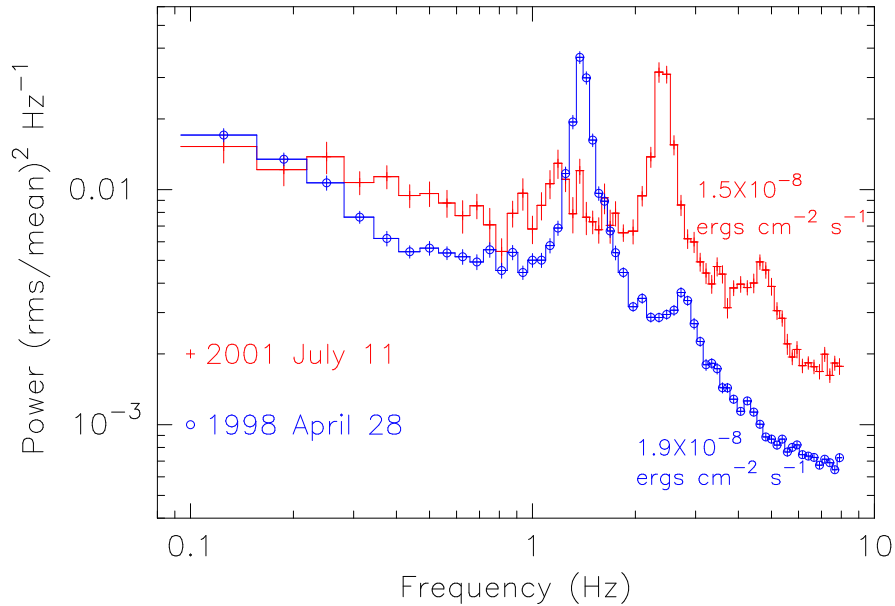}
\caption{Normalised power density spectra in 0.09 -- 9 Hz frequency
range observed on 1998 April 28 and 2001 July 11.  Strong QPOs are seen
at 1.4 Hz and 2.4 Hz in  PDS observed on 1998 April 28 and 2001 July 11
respectively. The total Compton scattered flux is also given. }
\label{qpos}
\end{center}
\end{figure}

In the last column of Table~\protect\ref{radio_sftab}, we give the 
calculated  absorption column density  N$_H$. The N$_H$ 
ranges from 10$\times$10$^{22}$ to 15$\times$10$^{22}$ cm$^{-2}$ 
which are higher than the commonly used N$_H$ $\sim$ 5 -- 6$\times$10$^{22}$ 
cm$^{-2}$ for spectral analysis of \grs.  The calculated N$_H$ shows
a tight correlation with the CompTT flux and not with the disk blackbody
flux. This  rules out the possibility that  the enhanced N$_H$ may be  
due to overestimation
of the disk normalisation. Yadav (2006) has discussed 
this in detail and compared these with  results obtained from Chandra and
ASCA data \cite{lee02,kota00}. Lee et al. (2002) 
have analysed Chandra \& RXTE X-ray data during a radio plateau state and 
have suggested presence of disk wind. Kotani et al. (2000) also came to
a similar conclusion using ASCA data.
The  wind terminal velocity is estimated to be of the order of
10$^8$ cm s$^{-1}$ ($\beta \le$ 0.01).
The lower limit of bolometric 
luminosity L$_{bol}$ $\sim$ L$_X$ 
= 6.4$\times$10$^{38}$ ergs s$^{-1}$ which is 0.35 of 
Eddington Luminosity  L$_{Edd}$, for a 
black hole of mass 14 M$_{\odot}$ \cite{lee02}. The corresponding lower 
limit of bolometric luminosity for the X-ray flux listed in 
Table~\protect\ref{radio_sftab} falls in the range  0.35 -- 0.48 of L$_{Edd}$.  
 When \grs is accreting near  
L$_{Edd}$, the presence of a radiation-driven wind is always expected and the 
wind density should be a strong function of the 
disk luminosity. 
  Our derived values of N$_H$ show strong dependence on observed total X-ray
  flux with correlation coefficient of 0.995  \cite{yada06a}.

The calculated power density spectra shown in Figure~\protect\ref{qpos} also 
independently lend support to the presence of wind during the plateau state.
 Figure~\protect\ref{qpos} shows PDS spectra observed on 2001 July 11
and 1998 April 28. The Compton scattered flux increases  from 
1.5$\times$10$^{-8}$ ergs cm$^{-2}$ s$^{-1}$ on 2001 July 11 to 
1.9$\times$10$^{-8}$ ergs cm$^{-2}$ s$^{-1}$ on 1998 April 28 while 
the observed
QPO frequency decreases from 2.4 Hz on 2001 July 11 to 1.4 Hz on 1998 April 28.
It is clear from Figure~\protect\ref{qpos} that the 
power at frequency $\nu > 0.2$ Hz is 
less in the PDS observed on   1998 April 28 than that observed  on  
2001 July 11. The fast
variability is suppressed by photon scattering in the enhanced  wind on
1998 April 28, hence reducing the power in the PDS at higher frequencies.
Shaposhnikov \& Titarchuk (2006) have discussed the decrease in the PDS power observed in 
Cyg X-1 at higher frequency ($\nu >$ 0.1 Hz) as the wind increases. In Cyg~X-3, 
the PDS power  in the low hard state  drops  to below 10$^{-3}$  
(rms/mean)$^2$ Hz$^{-1}$ at frequencies $\nu  > 0.1$ Hz as a dense wind
 from the companion always envelops the compact object \cite{yada06a}.

\begin{figure}[t]
\begin{center}
\includegraphics[angle=270,width=.7\textwidth ]{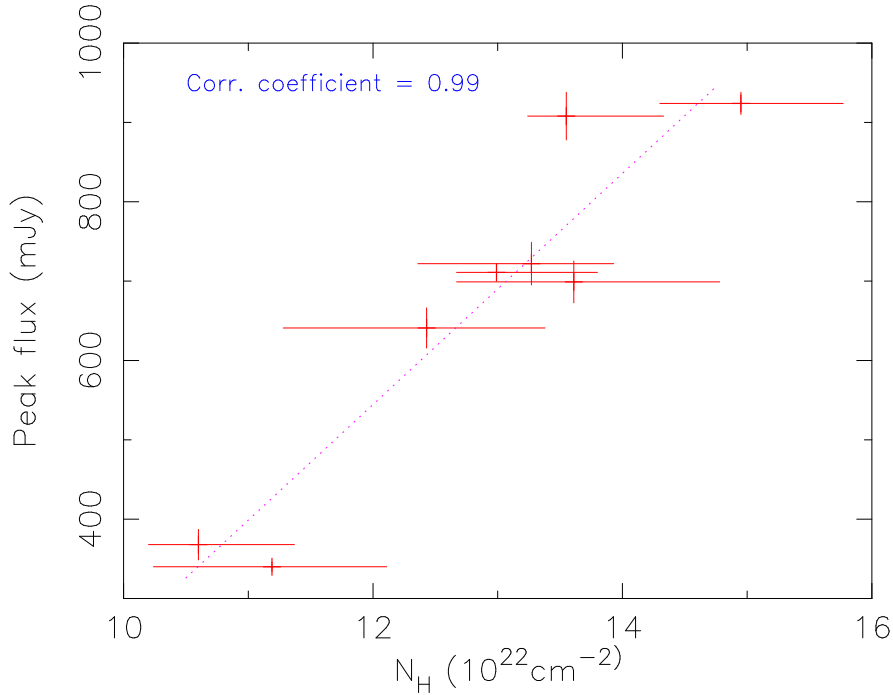}
\caption{Correlation between the  absorption column density, N$_H$ and 
the  peak flux of the superluminal radio
flares. The peak flux of a superluminal flare is calculated using exponential
profile fitting if there  is a gap in radio data.  The dotted line is a
linear fit to the data.}
\label{nhpeak}
\end{center}
\end{figure}

These results  support the internal shock model for the origin of superluminal flares 
\cite{kais00}.  The internal shock should form in the previously generated slowly moving
wind from the accretion disk with $\beta$ $\le$ 0.01  as the fast moving discrete
jet with $\beta$ $\sim$ 1 \cite{dhaw00} catches up and interacts with it. 
Both the components, 
 slow moving wind and  fast moving jet, are related to the accretion disk during  plateau state and 
the strength \& speed of these two components will determine the power of the internal shock.  The wind deposits a large amount of energy as $\dot{m}_{wind}$ approaches $\dot{m}_{accr}$ prior to the switch-on of  superluminal flare 
\cite{yada06}.  
Thus, the internal shock model can easily accommodate high  jet 
power requirement $\ge$ 10$^{38}$ ergs 
s$^{-1}$ \cite{fend99} and can explain the shifting from thick to thin radio emission during superluminal flares \cite{kais00}.
Our results in Figure~\protect\ref{nhpeak} which shows  the
peak flux of superluminal radio flares as a function of N$_H$   
strongly support our description of superluminal radio flares.
A fit to the data (dotted line)
 suggests that for wind strength corresponding to  
N$_H \le$ 8.3$\pm$1.5$\times$10$^{22}$ cm$^{-2}$, no superluminal jet
will  be produced. 
The absence of superluminal jets during
  the class $\beta$ in \grs is attributed to the absence of wind \cite{yada06}.
The calculated N$_H$ during class $\beta$ is below this  critical N$_H$ value
\cite{yada06a}.
 This model
  of superluminal jets can provide simple explanations for (1) why the 
  direction of superluminal jets may differ from the direction of the 
  compact jets, (2) why the phase lag is complex during plateaux, and
  (3) why compact (discrete) jets are absent in Cyg~X-3 \cite{yada06}.

\begin{figure}[t]
\begin{center}
\includegraphics[angle=0,width=0.6\textwidth ]{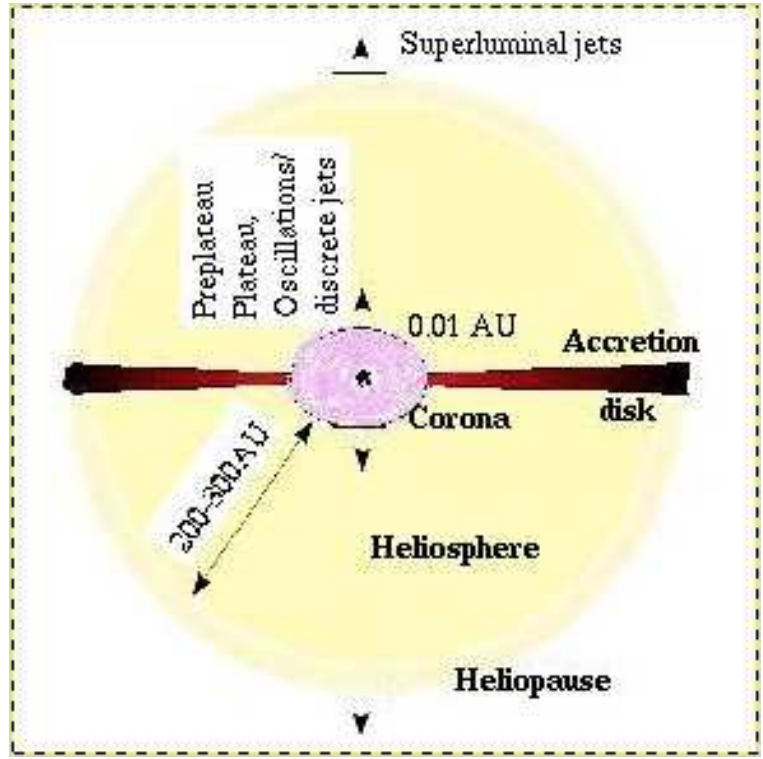}
\caption{Sketch diagram (not to the scale) of the heliosphere created by   
disk wind similar to the solar heliosphere. The locations are marked
where different types of radio jets are produced. Superluminal jets 
start at the boundary of the disk heliosphere while all other types of the
jets start somewhere close to the corona.}
\label{windsp}
\end{center}
\end{figure}

In Figure~\protect\ref{windsp}, we show a sketch of the heliosphere due to disk wind
and mark the locations of various types of radio emission
seen in \grs.     All the radio flares observed in \grs  can be
  broadly put into two groups on the basis of their flux, radio 
  spectrum and spatial distribution; (1) the superluminal  flares
  (200--1000 mJy) which have steep radio spectra and are seen at 
  large distances ($\ge$ 240 AU), and (2) all other flares (5--360 mJy) 
  which include the preplateau flares, radio oscillations \& 
  discrete flares  and the steady radio emission 
  during the plateaux. All these flares  have flat radio spectra
  and are observed close to the compact object. All
these radio flares are consistent with the ejection of an adiabatically 
expanding  self absorbing synchrotron  cloud from the accretion disk
and this cloud is supposed to consist of coronal mass. During the radio
plateau, it may be confined expansion due to the presence of
 a dense wind which agrees with the source size \cite{dhaw00}.
The IR \& radio oscillations (5--150)
  are periodic ejections of adiabatically expanding  self absorbing 
  synchrotron  clouds \cite{mira98,ishw02}. 
  The preplateau radio flares (50--360 mJy) are discrete ejections
  which are closely spaced in time and hence produce overlapped 
  radio flares. The preplateau
  flares have been modeled as  adiabatically expanding 
  self absorbing clouds ejected from the accretion disk 
  (like in the case of oscillations \& discrete jets) which 
  explain  reasonably well all the available observed data of
  time delay for radio emission at lower frequency \cite{ishw02}.
All these types of radio emission in \grs starts somewhere close
to the corona.

As discussed above, the  $\dot{m}_{accr}$, as inferred from the X-ray flux, 
is very high during the radio plateaux and
L$_{bol}$ approaches  L$_{Edd}$.
It is  suggested that such hot accretion disk during the radio plateaux  
always accompanies with a 
radiation-driven wind.
The N$_H$ is  tightly correlated  with the COMPTT flux. This is analogous to 
the solar wind originating from the dense solar corona. Since the disk wind
power \& speed are  higher than those of  solar wind (the average solar wind speed 
$\sim$ 450 km/s), 
the size of heliosphere
due to the disk wind may be around 200-300 AU 
(the solar heliosphere size is of 100-150
AU and it is supposed to vary with the solar activity). It is expected
that the size of disk heliosphere will increase as the wind power 
increases with 
disk luminosity as discussed earlier. As shown in Figure~\protect\ref{windsp},
superluminal radio jets are suggested to appear at the boundary of 
disk heliosphere. It
is expected that strong superluminal jets should appear at larger
distances than the distance where weak superluminal jets appear. 
A superluminal jet with peak flux of $\sim$ 200 mJy  was reported  
around 240 AU on 2001 July 16\cite{dhaw03}  while strong superluminal jets 
are suggested to appear 
around 500 AU (the peaks of these superluminal flares are
not observed  but the data are  consistent with the above suggestion) \cite{dhaw00}.
This can be easily tested in future with VLBA and other high resolution
radio data if we catch the start of 
superluminal flares in \grs as well as in other LMXBs.
\section{Conclusions}
	We have  provided a tight
correlation between accretion disk and superluminal jet parameters. We
find  that $\dot{m}_{accr}$ 
during the plateaux is very high and suggest 
that the accretion disk during the radio plateaux  is always associated with a 
radiation-driven  wind. The internal shock  forms in the previously generated
slowly moving wind (during plateau)  with $\beta$ $\le$ 0.01  as the
fast moving discrete jet (usually at the end of plateau) with $\beta$ $\sim$ 1
catches up and interacts with it.
 The strength \& speed of these two components  determine the power of the 
internal shock as well as of the superluminal jets.
The peak flux of the superluminal jets is
tightly  correlated with the wind power. We discuss the implications of this
work and some of these can be checked in future.

\noindent {{\bf Acknowledgment:}
\it {The author  thanks  the RXTE PCA/HEXTE and NSF-NRAO-NASA Green Bank Interferometer 
teams for making 
their data publicly available.  The Green Bank Interferometer is a facility 
of the National Science Foundation operated by the NRAO in support of NASA 
High Energy Astrophysics programs.}}

\end{document}